\documentclass[aps,prd,10pt,nofootinbib,twocolumn,eqsecnum,showpacs,showkeys,superscriptaddress,preprintnumbers,altaffilletter]{revtex4-1}
\DeclareUnicodeCharacter{0327}{\c{}}
\usepackage[utf8]{inputenc}

\usepackage{graphicx}
\usepackage{amsmath}
\usepackage{multirow}
\usepackage{hyperref}
\usepackage{soul}
\usepackage{graphicx}
\usepackage{dcolumn}
\usepackage{amssymb}
\usepackage{amsfonts}
\usepackage{amsbsy}
\usepackage{color}
\usepackage{rotating}
\usepackage[english]{babel}
\usepackage{multirow}
\usepackage{float}

\newcommand{\be}{\begin{equation}}
\newcommand{\ee}{\end{equation}}
\newcommand{\bea}{\begin{eqnarray}}
\newcommand{\eea}{\end{eqnarray}}

\hypersetup{
    colorlinks=true,
    linkcolor=blue,
    filecolor=magenta,
    urlcolor=cyan,
    citecolor=cyan
}

\begin{document}
\title{Mass$-$to$-$Horizon Relation and Entropy Beyond the Bekenstein–Hawking Limit}
\author{Hussain Gohar}
\email{hussain.gohar@usz.edu.pl}
\affiliation{Institute of Physics, University of Szczecin, Wielkopolska 15, 70-451 Szczecin, Poland}

\date{\today}

\begin{abstract}
\noindent
We introduce a generalized mass–horizon relation applicable to cosmological horizons. This formulation provides a unified framework for deriving a broad class of Bekenstein entropy extensions motivated by statistical mechanics, quantum gravity, and phenomenological considerations, through the application of the Clausius relation together with the Hawking temperature.
We further introduce this notion as a foundational framework for constructing generalized entropy forms that remain consistent with thermodynamic laws and the holographic principle.
\end{abstract}
\maketitle
\textbf{\textit{Introduction--}}
The entropy of black holes \cite{Bekenstein:1973ur} is fundamentally supported by Hawking's area theorem \cite{Hawking:1971vc}, which has recently been validated by gravitational wave data \cite{KAGRA:2025oiz}. This theorem asserts that the horizon area of a black hole does not diminish, providing an analogy with the second law of thermodynamics. As a result, black hole entropy $S_{bh}=\frac{k_Bc^3A}{4\hbar G}$ is quantitatively expressed by the area of the horizon $A$, employing $k_B$ as the Boltzmann constant, $c$ as the speed of light, $G$ as Newton's gravitational constant, and $\hbar$ as Planck's reduced constant. This formulation is further reinforced by Hawking's quantum evaporation process \cite{Hawking:1974rv}, and the corresponding Hawking temperature for $S_{bh}$ is deduced within the framework of quantum field theory, represented as $T_h = \frac{\hbar \, \kappa}{2 \pi k_B c}$, where $\kappa$ is identified as the surface gravity on either black hole or cosmological horizons.
Numerous extensions of $S_{bh}$ have been formulated, rooted in diverse principles of thermodynamics, statistical mechanics, and quantum gravity, alongside other phenomenological frameworks
\cite{Rovelli:1996dv, Meissner:2004ju, Medved:2004yu, Das:2007mj, Barrow:2020tzx,Tsallis:2012js,Czinner:2015eyk,Alonso-Serrano:2020hpb,Kaniadakis:2005zk,Gohar:2023lta,Nojiri:2022aof,Nojiri:2022dkr}. Moreover, a wide range of applications \cite{Jacobson:1995ab,Alonso-Serrano:2020dcz,Cai:2005ra,Padmanabhan:2003pk,Li:2004rb, Verlinde:2010hp, Komatsu:2013qia} in gravitational and cosmological contexts have been developed through the use of $S_{bh}$ and various generalized definitions of $S_{bh}$, in conjunction with the Hawking temperature $T_h$. In the majority of these applications, the holographic principle \cite{tHooft:1993dmi,Susskind:1994vu} is predominantly utilized, with the defined quantities of mass/energy ($M/E$), entropy ($S$), and temperature ($T$) being associated with chosen holographic horizon $L$ in accordance with the Clausius relation, $dE=TdS$, and the first law of thermodynamics.

A significant but often overlooked relation in holographic applications in the context of cosmological applications, is the linear {\it mass-to-horizon relation} (MHL), 
$M=\gamma \, \frac{c^2}{G}L$, 
where $\gamma>0$ is a dimensionless free parameter.
 This relation is implicitly assumed in cosmological applications while applying Clausius relation and the first law of thermodynamics. For the first time, this assumption is explicitly introduced by us in \cite{Gohar:2023hnb} and a generalized version in \cite{Gohar:2023lta}.
In the case of a Schwarzschild black hole, the aforementioned linear MHL is delineated as $M=\frac{c^2}{2G}r_{+}$, where $r_{+}$ denotes the radius of the event horizon. While this linear relation is apparent for the Schwarzschild black hole, in cosmological holographic scenarios, it remains an assumption, albeit crucial, for maintaining thermodynamic consistency. Thermodynamic consistency in this context denotes that any $S$ and $T$ associated with holographic horizon, when employed in $dE=TdS$, must ensure the equivalence of energy $E$ and mass $M$.
Furthermore, the holographic principle requires that the entropy, mass, and temperature associated with the chosen horizon must be thermodynamically consistent. 
The generalizations beyond $S_{bh}$, the emphasis is placed on extending $S_{bh}$ while maintaining the integrity of $T_h$. The inquiry posed is whether the modifications of $S_{bh}$ in combination with $T_h$ yield thermodynamic consistency \cite{Cimdiker:2022ics}. Furthermore, the inquiry extends to whether $T_h$ is applicable in conjunction with these generalized entropies.
In response to these inquiries, we propose an innovative notion of generalized mass-to-horizon relation, ensuring that thermodynamic consistency is preserved across various extensions of $S_{bh}$ when used with $T_h$.

\textbf{\textit{Generalized Mass-to-Horizon Relation--}} 
We propose the following generalized mass–to-horizon relation, expressed as
\[
M = \gamma \, \frac{c^2}{G}\ell_{\rm Pl} 
\left[ \frac{L}{\ell_{\rm Pl}} \;\mp\; \beta \left(\frac{L}{\ell_{\rm Pl}}\right)^{3-\alpha} \right]^m,
\]
where
\(\beta > 0\) is a dimensionless free parameter,
\(m > 0\), \(\alpha \in \mathbb{R}\) are dimensionless exponents, and 
\(\ell_{\text{Pl}} = \sqrt{\hbar G / c^3}\) is the Planck length. 
With this generalized MHL, the Clausius relation ensures that \(T_h\) and all associated extensions of \(S_{bh}\) remain thermodynamically consistent with the holographic principle. 

For instance, setting \(\beta=0\) and \(m=2\delta-1\) reproduces the Tsallis-Cirto entropy characterized by the nonextensive parameter \(\delta\), maintaining holographic consistency with \(T_h=\frac{\hbar c}{2\pi k_B L}\) when \(L\) represents the apparent or Hubble horizon in cosmological scenarios. Likewise, choosing \(m=1+\Delta\) yields the Barrow entropy, where \(\Delta\) reflects the Barrow parameter associated with the quantum fractal structure of the horizon. 
Of particular significance, in the limits \(m=1\) and \(\alpha \rightarrow 4\), the generalized MHL ensures thermodynamic consistency between \(T_h\) and the quantum-gravity–inspired extensions of \(S_{bh}\). This same limit also recovers entropy corrections due to quantum entanglement of fields across the horizon. 

Furthermore, employing the Clausius relation with \(T_h\) and assuming \(\beta \ll 1\) leads to the most generalized form of entropy,
\[
S_G = 2\pi k_B\gamma \left[\frac{m}{m+1}
   \left( \frac{L}{\ell_{\rm Pl}} \right)^{m+1}
  \mp  m\frac{\sigma - 1}{\sigma}\beta
  \left( \frac{L}{\ell_{\rm Pl}} \right)^{\sigma}
\right],
\]
where \(\sigma = m + 3 - \alpha\).
For \(m=\gamma=1\) and \(\beta=0\), it simplifies to \(S_{bh}\); in the limit \(m=1\) and \(\alpha \to 4\), it yields quantum-gravity corrections to \(S_{bh}\)~\cite{Rovelli:1996dv, Meissner:2004ju, Medved:2004yu}.
Similarly, for \(m=1\), it captures entanglement-induced entropy modifications~\cite{Das:2007mj}, while setting \(1+m=2\delta\) and \(\beta=0\) recovers the nonextensive Tsallis-Cirto entropy~\cite{Tsallis:2012js}.
The Barrow entropy~\cite{Barrow:2020tzx} is obtained for \(m=1+\Delta\) with \(0 \leq \Delta \leq 1\), implying \(1 \leq m \leq 2\) is compatible. 
In the \(\beta \neq 0\), \(\alpha \rightarrow 4\) limit, the framework further yields—for the first time—quantum-gravity–induced corrections to both Tsallis-Cirto and Barrow entropies.

\textbf{\textit{Discussion--}}
In this letter, we present a rigorous thermodynamic framework grounded in the holographic principle to derive different extensions of $S_{bh}$, ensuring the preservation of thermodynamic and holographic consistency through our newly formulated generalized {\it mass-to-horizon relation}. While we propose a novel and consistent methodology to modify $S_{bh}$, focusing on the most important generalized versions of $S_{bh}$ prevalent in cosmological applications, this methodology can be equally applicable to other phenomenological approaches to introduce new or alternative generalized versions of $S_{bh}$.

Geometrically, the notion of MHL is vital for the computation of geometrical thermodynamic quantities. Notably, for $m=1$, $\beta=0$, and $\gamma=1/2$, it delineates the Misner-Sharp mass/energy for a scenario exhibiting spherical symmetry. Additionally, the geometric representation of the Clausius relation at the apparent horizon is explored through a mass-like function in \cite{Gong:2007md}. Therefore, whether considering the geometric interpretation of the Clausius relation or its conventional thermodynamic counterpart, the novel notion of MHL is fundamentally important for applications of the gravity-thermodynamic correspondence in general and, particularly, in cosmological scenarios. 

Furthermore, the integration of the geometrical gravity theory with each generalized version of $S_{bh}$ entropy via the approach given in \cite{Jacobson:1995ab,Alonso-Serrano:2020dcz} to derive modified field equations becomes essential for elucidating the geometric origins of these extensions of $S_{bh}$. For instance, as shown in \cite{Lu:2024ppa} for the Tsallis-Cirto entropy case, such generalized entropy results in a theory with a varying effective gravitational constant, in which $G_{\rm eff}$ is contingent upon the horizon area. In contrast, the Barrow entropy indicates that the cosmological constant is recalibrated in terms of the horizon area \cite{DiGennaro:2022grw}, although the field equations remain unchanged for both scenarios. Therefore, the exploration of modified gravitational frameworks pertinent to each form of entropy is of significant importance, albeit challenging, especially when considering diverse extensions of $S_{bh}$. 

Moreover, it is a well-established fact that the Hawking temperature $T_h$ is independent of theoretical frameworks and is consistently formulated using quantum field theory. One might inquire whether employing a linear MHL alongside various extensions of $S_{bh}$ permits modifications to $T_h$ via the Clausius relation while still retaining a consistent mass-energy relationship. However, justifying a modified $T_h$ within the framework of quantum field theory proves challenging. Our recent findings \cite{Gohar:2023hnb} indicate that, when we start with generalized entropies with corresponding modified $T_h$, we end up with the same evolution equations. For example, within entropic cosmological models, {\it we have demonstrated that, with consistent thermodynamic quantities defined on the Hubble horizon, satisfying the Clausius relation, and employing a linear MHL, all the investigated nonextensive entropic force models are identical to the original entropic force models derived from the standard Bekenstein entropy and the Hawking temperature.} This suggests that, in addition to the absence of justification for a modified $T_h$ within the framework of quantum field theory, distinguishing the modifications to the applications of gravity-thermodynamic correspondence in cosmology, arising from various concepts of generalized entropies, presents a significant challenge. Consequently, the incorporation of the mass-to-horizon relation, in conjunction with all generalized entropies and their consistent thermodynamic association with the Hawking temperature, constitutes a more effective heuristic framework for implementation in the realms of gravity and thermodynamic applications.

\textbf{\textit{ Conclusion--}} In conclusion, any attempt to generalize black hole entropy beyond the Bekenstein framework should be approached with extreme caution due to our incomplete understanding of standard Bekenstein entropy. 
The extension of Bekenstein entropy across multiple parameters, as performed in \cite{Nojiri:2022aof,Nojiri:2022dkr}, simply because it encapsulates various specific entropies under different parameter settings, is not deemed a methodologically rigorous approach. Consequently, this letter concentrates on the thermodynamic consistency of quantities associated with cosmological horizons within holographic scenarios. For this purpose, our newly introduced generalized mass-to-horizon relation plays a crucial role. It is worthwhile to engage in similar investigations for other generalized entropies, such as R\'enyi or Kaniadakis entropies, or any novel entropy that might be derived from distinct modifications of MHL. Moreover, these considerations can be extended to black hole horizons, including those of charged and rotating black holes. Ultimately, the implementation of gravity-thermodynamic correspondence necessitates a cautious approach to mitigate potential thermodynamic inconsistencies, especially considering the lack of a foundational framework for the application of generalized entropies within the contexts of gravity and cosmology. In summary, this methodology holds the promise of offering innovative thermodynamic insights into the study of gravity, while simultaneously providing a platform to reassess previous studies documented in the existing literature..

\bibliographystyle{apsrev4-1}
\bibliography{ref}

\end{document}